\documentclass{article}

\usepackage{arxiv}
\usepackage[utf8]{inputenc} 
\usepackage[T1]{fontenc}    
\usepackage{hyperref}       
\usepackage{url}            
\usepackage{booktabs}       
\usepackage{amsfonts}       
\usepackage{nicefrac}       
\usepackage{microtype}      
\usepackage{lipsum}
\usepackage{graphicx}
\usepackage{amsmath} 
\usepackage{tabularray}
\usepackage[table]{xcolor}
\usepackage{caption}
\usepackage{cite}  

\graphicspath{ {./images/} }

\title{GATE: Adaptive Learning with Working Memory by Information Gating in Multi-lamellar Hippocampal Formation}

\author{
 Yuechen Liu, Zishun Wang, Chen Qiao$^*$ and Zongben Xu\\ \\
  School of Mathematics and Statistics\\
  Xi'an Jiaotong University\\
  \texttt{yuechenliu,wangzs11611@stu.xjtu.edu.cn}\\ \texttt{qiaochen,zbxu@mail.xjtu.edu.cn} 
}

\begin{document}
\maketitle
\date{}

\begin{abstract}

Hippocampal formation (HF) can rapidly adapt to varied environments and build flexible working memory (WM). To mirror the HF's mechanism on generalization and WM, we propose a model named Generalization and Associative Temporary Encoding (GATE), which deploys a 3-D multi-lamellar dorsoventral (DV) architecture, and learns to build up internally representation from externally driven information layer-wisely. In each lamella, regions of HF: EC3-CA1-EC5-EC3 forms a re-entrant loop that discriminately maintains information by EC3 persistent activity, and selectively readouts the retained information by CA1 neurons. CA3 and EC5 further provides gating function that controls these processes. After learning complex WM tasks, GATE forms neuron representations that align with experimental records, including splitter, lap, evidence, trace, delay-active cells, as well as conventional place cells. Crucially, DV architecture in GATE also captures information, range from detailed to abstract, which enables a rapid generalization ability when cue, environment or task changes, with learned representations inherited. GATE promises a viable framework for understanding the HF's flexible memory mechanisms and for progressively developing brain-inspired intelligent systems.
\end{abstract}


\section{Introduction}
When facing a new task or unfamiliar environment, an agent must firstly decide which new information (e.g., observations or self-generated behaviors) to keep, depending on its past experiences and the potential relevance of the information for future use. Subsequently, the agent should keep this important information, and adapt to a new task based on the cross-time associations between information and the task. For instance, in a listening comprehension exercise, the students need to utilize their prior practice experience, and discern essential information to retain. Next, the information should be “kept in mind” as they listen, allowing them to answer subsequent questions accurately. The associations formed between retained information and the correct answers given afterward enhance students' performance in future exercises.

Working memory (WM, also known as temporal dependency learning), precisely corresponds to the ability to hold a limited amount of information in mind temporarily, acting as a substrate for predicting, planning, association forming, etc. \cite{baddeley12}. Simultaneously, generalization, or adaptability, is the capacity to swiftly understand new task rules or new environment based on past experience \cite{banich11}. Both WM and generalization play crucial roles in complex cognitive processes and they interact closely, yet the underlying mechanism supporting these abilities in mammal, and the ways in which they interact, are still uncovered. Meanwhile, generalization requires the agent not only acquires detailed knowledge of the novel environment, but also attains abstract understanding of the task, selectively disregarding irrelevant details. However, how WM can retain these two seemingly contradictory aspects, is still unknown. 

\begin{figure}   
\centering
\includegraphics[width=0.9\textwidth]{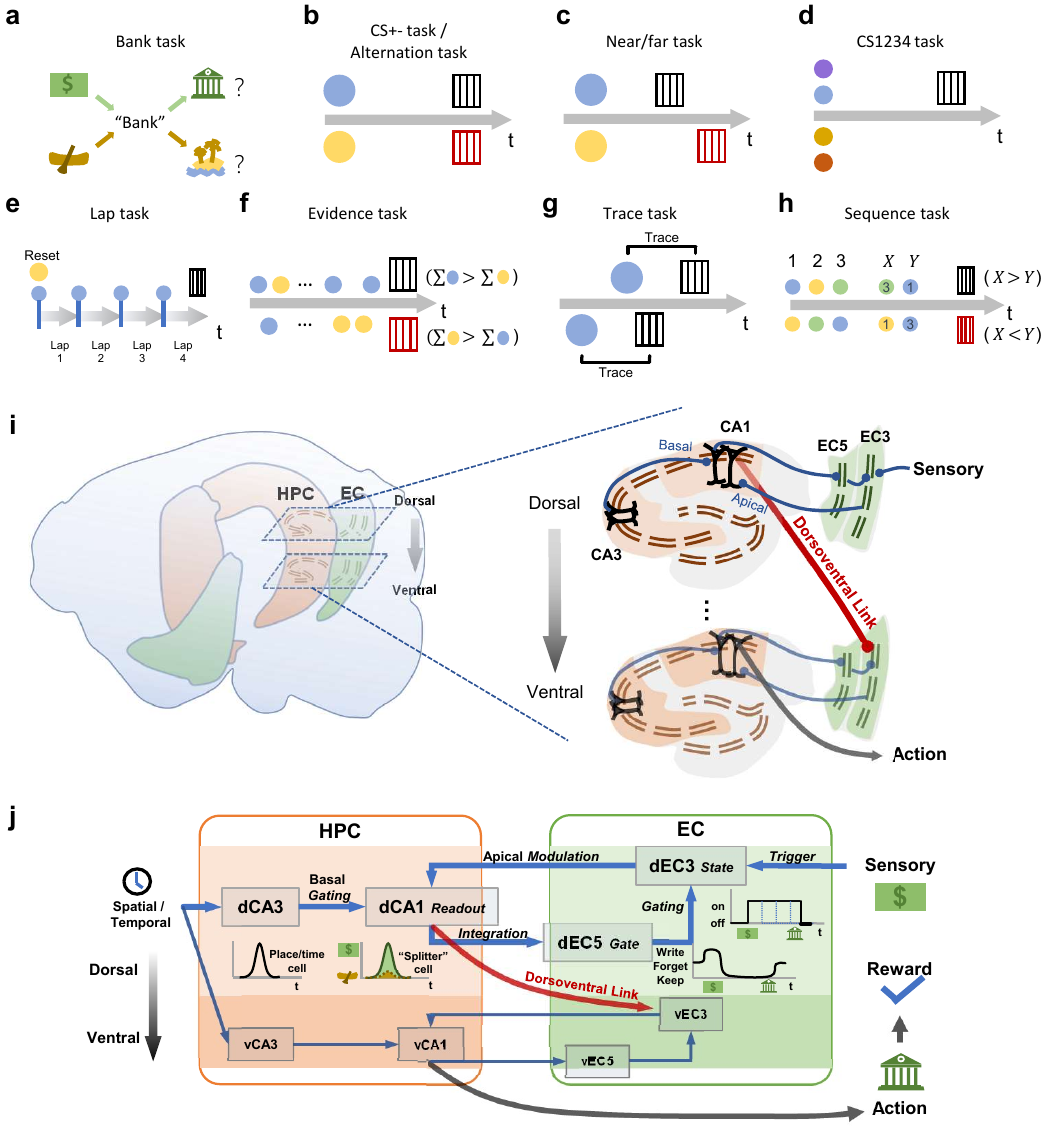}
\caption{
\textbf{Working memory tasks, hippocampus formation structure and GATE model. }
(\textbf{a}) Semantic paradigm of working memory tasks in language understanding. To determine whether the word “Bank” refers to financial bank (indicated by cash) or river bank (indicated by boat rolling), one needs to keep the context in mind.
(\textbf{b-h}) Task descriptions.
(\textbf{b}) CS+- task: Two cues are presented (randomly one per trial), each indicating a specific correct choice at the track's end.
(\textbf{c}) Near/far task: Similar to CS+ but requires actions at different locations.
(\textbf{d}) CS1234 task: Two of four cues are actionable; the others are not.
(\textbf{e}) Lap task: The agent resets, completes four laps, and acts at the end of the fourth lap. The environment remains unchanged across laps.
(\textbf{f}) Evidence task: The agent identifies which of two cues occurs more frequently in a Poisson sequence.
(\textbf{g}) Trace task: The agent acts after a fixed delay following a random cue.
(\textbf{h}) Sequence task: The agent determines which of two repeated cues appeared earlier in a three-cue sequence.
(\textbf{i}) Semantic HF connectivity. Hippocampus (HPC) and EC form a re-entrant loop: EC3 → CA1 → EC5 → EC3. CA3 and EC3 inputs dominate CA1 basal and apical dendrites, respectively. Adjacent lamellas connect dorsoventrally. Sensory input drives dorsal EC3, and ventral CA1 outputs actions.
(\textbf{j}) Work flow of GATE in Bank task. EC3 processes sensory input (e.g., “Cash”) and modulates CA1 readout (e.g., activates after “Cash”, not “Boat”). CA3 gates CA1 timing; EC5 integrates CA1 signals and regulates EC3 memory states (write, retain, erase). Correct predictions (e.g., “Financial Bank”) yield rewards.
} 
\label{fig1}
\end{figure}

The hippocampal formation (HF), which plays a crucial role in both WM and generalization, offers constructive insights. WM shows significant relationship with HF \cite{daume24}, both on neural representation and network structural aspect. For neural representation aspect, HF forms a state map, known as the cognitive map, which reads out task-relevant information temporarily kept by WM \cite{biane_23,pastalkova_08,sun_23,wang15} (Fig. \ref{fig1}a-h). Further, researches have also shown that HF learn to compose the maintained information in the cognitive map, e.g., the stimuli order \cite{fortin_02}, evidence accumulation \cite{nieh_21}, lap number \cite{sun_20}, or time point \cite{fortin_02,macdonald_11,mcechron_97} of the cue sequence. Although numerous researches uncover that the cognitive map plays an important role in WM, the mechanism remains uncertain. Meanwhile, persistent activity in entorhinal cortex (EC) \cite{tahvildari_07,jochems13} offers a direct way of memory maintenance, acting as a Markov-chain in behaving rodents \cite{grienberger_22}. The activity also displays dependency on cues or rewards, decays thereafter, which may explain how writing and forgetting occur, but how keeping and reading out occur are still vague. At the structural level, EC3-CA1-EC5-EC3 recurrence (also called big-loop in \cite{koster_18}) on transversal axis of HF, mediates the cross-time association process (Fig. \ref{fig1}i). 

On the other hand, mammal adaptability to new task also strongly relies on functional HF \cite{rubin14}. HF offers a flexible cognitive map, i.e., the cognitive map swiftly remaps when the environment or task changes \cite{tanila_97,colgin08}. Along the longitudinal axis (dorsoventral axis, or DV axis), HF shows a role in generalization. During learning, CA1 acts differently in dorsal part and ventral part \cite{fanselow10}. Specifically, detailed representations of environmental stimuli in the dorsal CA1 (dCA1) form at once when facing new situation. After a short period of learning, the detailed dCA1 encoding results are transferred to more abstract, task-relevant representations in ventral CA1 (vCA1) that predict salient outcomes (e.g., rewards and punishments) \cite{biane_23}. In this sense, dCA1 offers highly detailed observations of the environment, while vCA1 condenses and synthesizes the information from these observations, making it easier for the agent to generalize. 

Although many properties of HF have been revealed, the mechanisms of HF contributing to WM and generalization remain uncertain. To address these questions, several learning models have been proposed, including the Tolman-Eichenbaum machine (TEM) \cite{whittington_20}, Hebbian-RNN \cite{kappel_14}, clone structured cognitive graph (CSCG) \cite{george_21}, and plasticity model \cite{cone_24}. However, it is worth noting that, the vital role of HF in adapting to new environment or new task have not been taken into consideration. 

To model how WM and generalization are formed and how they are integrated in HF, we propose a network model named generalization and associative temporary encoding (GATE) (Fig. \ref{fig1}j). GATE firstly employees an EC3 population persistent activation model, which receives and maintains task variables. Next, re-entrant loop structure is introduced into GATE, where EC3 acts as an information container, CA1 acts as the information reader, CA3 acts as the temporal or positional gate, and EC5 acts as an attentional integrator. In this model, information kept by EC3 population could be selectively read out by CA1 when CA3 gate is open, integrated in EC5 and then transferred to EC3, gating information in EC3 in a self-controlling manner. Finally, GATE makes use of both the transversal axis along with longitudinal axis, allowing it to learn non-trivial tasks via forming compound and abstract representations, and igniting its adaptability as well. 

As a result, GATE achieves excellent performance in multiple tasks, which cover several types of WM tasks including information maintaining \cite{biane_23,sun_23,ainge_07} and composing  \cite{biane_23,fortin_02,nieh_21,sun_20,grienberger_22}. During learning, GATE not only develops task-relevant representations comparable with recent researches, including splitter cell \cite{zhao_22} (or alternative cell \cite{ainge_07}), trace cell \cite{mcechron_97}, evidence cell \cite{nieh_21} and lap cell \cite{sun_20}; but also captures faithful external information in dorsal part, while information with significant meanings is captured in ventral part, just as the rodent HF does \cite{biane_23}. These results validate GATE consists with biological mechanism. Moreover, the learnt cognitive map is inherited during generalization, and GATE learns faster and faster, mirroring the adaptability of HF \cite{sun_23}. Finally, several experimentally verifiable predictions can be derived based on GATE.

\section{Result}
\label{sec:headings}

\subsection{EC3 Persistent activity learning: A trainable random population model with working memory functionality}

To build a model with WM functionality, we first utilize EC3's Persistent activity \cite{tahvildari_07} to accomplish the information maintaining task, which requires the agent to retain the external-driven stimulus (e.g., CS+- task, Near/far task). Roughly, four step will be taken, write down important information at first, keep it, then read it out when needed, and forget it when it is not useful anymore.

EC3 persistent activity shows the potential in write, keep and forget. Firstly, a part of EC3 cue-tuning neurons, responds to nearly any landmark, but the response intensity varies across different landmarks, forming an encoding of external information \cite{kinkhabwala_20}. Secondly, EC3 have been shown to function like on-off switches, with the firing rate being all-or-nothing, resembling a Markov chain. Although EC3 shows strong randomness, it encodes task-relevant information, including cue and reward locations, even at the early stage of learning \cite{grienberger_22}.

These properties of EC3 persistent activity encourages us to build a population-level model to describe the ratio of “on cell” in EC3 subgroups, $r(t)$. It changes rely on EC3 input $I(t)$, via two independent ways: on-off transition with probability $p_{10}(I)$, and off-on transition with probability $p_{01}(I)$ (Fig. \ref{fig2}a-b). $r(t)$ forms a one-order ODE, which converges towards the global asymptotic stable point $r_\infty$, according to the time constant $\tau$, if $I$ is fixed (see Section \ref{Methods}). As $I$ increases from low to high, the model sequentially exhibits different stages of keep (with large $\tau$), forget (with moderate $\tau$ and $r_\infty$), and write (with low $\tau$ and large $r_\infty$) (Fig. \ref{fig2}d). As a result, the model shows activation and maintenance when stimulus inputs are given, similar with experimental results in \cite{grienberger_22} (Fig. \ref{fig2}b,e). Additionally, compared to discrete Markov chain, our model can be trained using back-propagation.

\subsection{Re-entrant loop: A self-controlling information gating model with single-lamellar structure}

The EC3 population model implements the write, keep, and forget functions. Moving forward, it is necessary to explore how to design the model to incorporate a selectively read function and enable self-regulation of its input $i$. 

\begin{figure}
\centering
\includegraphics[width=1\textwidth]{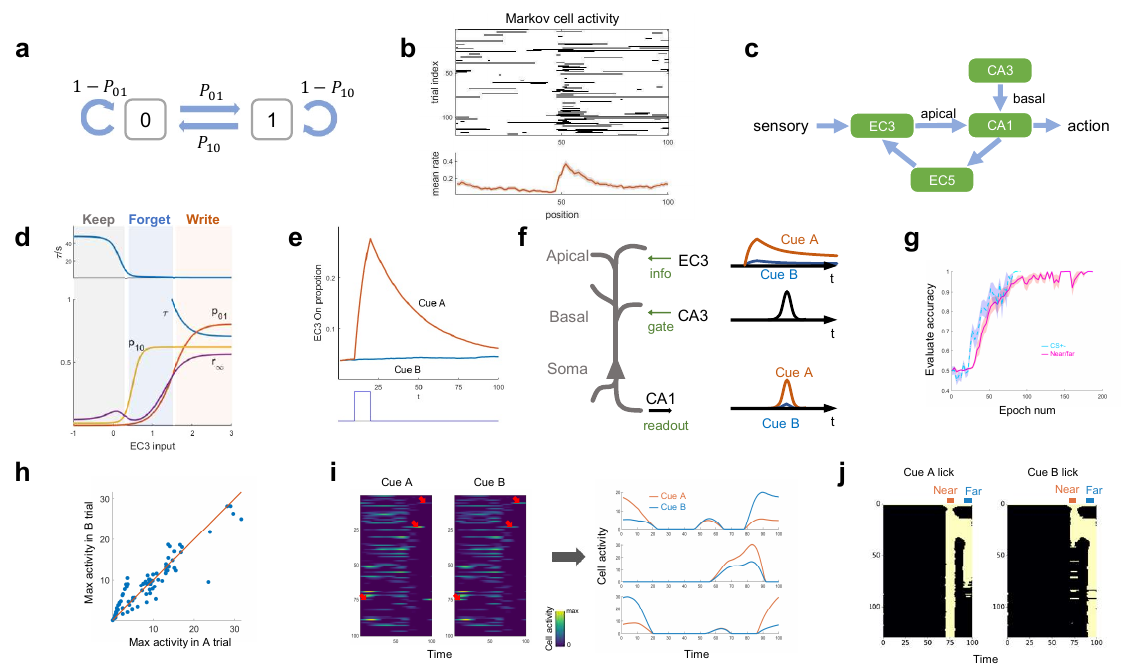}
\caption{
\textbf{Single-lamellar model learns to maintain information.}
(\textbf{a}) EC3 0/1 state transition. 
(\textbf{b}) EC3 neuron shows Markov chain property. Upper, activity of a simulated EC3 neuron in response to a pulse input at (45, 55), black indicates on state. Lower, mean neuron rate across trials (shadow=SEM), similar to \cite{grienberger_22}.
(\textbf{c}) Working flow of single-lamellar model, forming a re-entrant Loop.
(\textbf{d}) EC3 output governed by $P_{01}$, $P_{10}$, $r_\infty$, $\tau$ basing on EC3 input. Shadowed areas highlight stages of information processing.
(\textbf{e}) Sensory stimulus generate different EC3 activity. Upper, mean rate of EC3 subgroup. Lower, sensory stimulus in cue A trial (blue) and B trial (gray).
(\textbf{f}) CA1 neuron model. Left, semantic CA1 neuron structure. Right, semantic input and output of CA1 neuron. EC3 drives CA1 apical tuft, CA3 drives CA1 basal dendrites, creating a CA1 Splitter cell \cite{zhao_22}.
(\textbf{g}) Training accuracy of Near/far task (shadow=SEM).
(\textbf{h}) The model learns both conventional place cell and splitter cell. Red line indicates x=y. 
(\textbf{i}) CA1 activity by trial type. Left, firing rate, red arrows indicates representative cells shown on right, which shows similar responds to \cite{zhao_22}.
(\textbf{j}) Agent actions during Near/far task training. Bright area highlights “lick” action. 
}
\label{fig2}
\end{figure}

Since EC3 input towards CA1 distal (or, apical) dendrites are severely attenuated on their way to CA1 soma, their ability to ignite somatic spikes is poor, but can be facilitated when EC3 input is paired with modest CA3 input on basal dendrites of CA1 (also known as gating mechanism) \cite{jarsky_05}. Considering CA3 can encode both location and time \cite{salz_16}, the gating mechanism prompts us to view CA3 as regulating CA1 “when” or “where” to readout information, which is retained and provided by EC3  \cite{manns_05} (Fig. \ref{fig2}f). Furthermore, the readout CA1 activation can thereafter determine behavioral actions through a linear transformation. 

EC5 is an ideal candidate to regulate the EC3 input. EC5 also shows persistent activity, but act as a numerical integrator \cite{egorov_02}, i.e., EC5 changes its persistent firing rate only when strong excitatory or inhibitory input is given. Combined with HF structure, we introduce a EC3-CA1-EC5-EC3 re-entrant loop in our model, i.e., information from EC3 is selectively readout by CA1, integrated in EC5, then determines the stage of each EC3 subgroups in next time step – whether to write down new information, hold previous one or forget it. 

Re-entrant loop structure inspires the single-lamellar model. The input of this model is the sensory encoding through EC3 and positional or temporal encoding through CA3, while output is the current behavior decision (Fig. \ref{fig2}c). The model preforms well in information maintaining tasks, including CS+- and Near/far task (Fig. \ref{fig2}g). Furthermore, the single-lamellar model develops a cellular level task-relevant representation (also known as splitter cell \cite{ainge_07,zhao_22}) covering nearly the whole track (Fig. \ref{fig2}h,i), which can directly guide the agent to choose correct decision. Meanwhile, there are still conventional CA1 place cells without trail-type tuning (non-splitter cells), which plays a role in guiding EC5 “where” or “when” to adjust the EC3 input, or guiding the agent to do what. In short, the single-lamellar model enables CA1 to selectively readout information held by EC3 populations, then guide the behavior decision; but also enables EC5 to integrate and guide the information maintaining process in EC3. Additionally, the single-lamellar model also agent learns the task strategies step-by-step as rodents do (Fig. \ref{fig2}j). The agent licks over the track at beginning of training, then lick at both reward site (Near and Far) after a short while of training, and eventually develops an optimal lick behavior at the correct reward site \cite{sun_23}.

\subsection{Dorsoventral axis based multi-lamellar architecture: Learning externally and internally driven information}

For complex tasks in real world, merely retaining details from the environment (externally-driven information) is insufficient. Rather, the agent should also extract and compose information that fit for the task, from the numerous details kept in mind. Specifically, many tasks possess the identical cue type but lead to different results, depending on the order \cite{fortin_02}, accumulation \cite{nieh_21,sun_20}, or time \cite{macdonald_11,mcechron_97} of the cue delivered (Fig. \ref{fig1}e-h). In other words, it is important to condense the redundant observations into task-relevant abstract information (internally-driven information) during the learning process. HF offers inspiration that externally driven information is encoded in dCA1 while internally driven information in vCA1.

\begin{figure}
\centering
\includegraphics[width=1\textwidth]{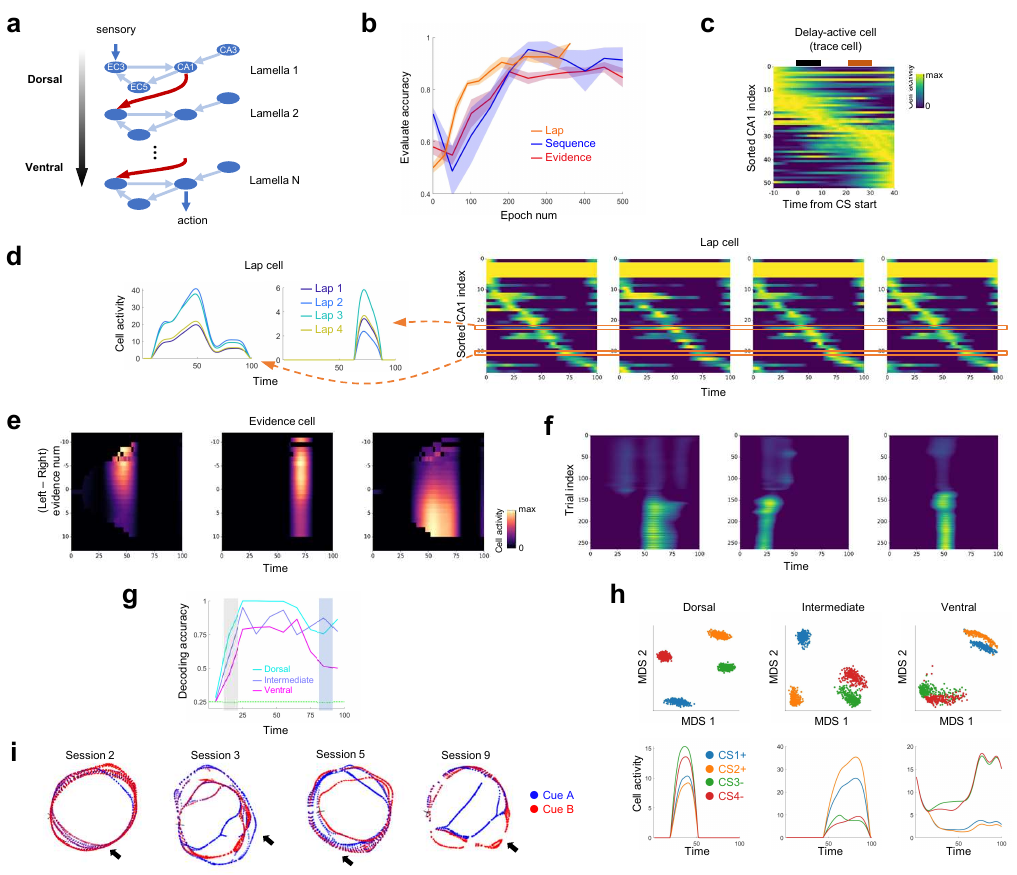}
\caption{
\textbf{Multi-lamellar model learns complex working memory tasks. }
(\textbf{a}) Semantic paradigm of multi-lamellar model.
(\textbf{b}) Evaluation accuracy in task Lap, Sequence, and Evidence. Shadow area indicates SEM.
(\textbf{c}) Delay-active cells (trace cells) in Trace task, sorted by max activity location, similar to \cite{masuda_20} . CS, conditional stimulus zone (black bar); US, unconditional stimulus zone (brown bar). 
(\textbf{d}) Cell representation in Lap task. Orange boxes and dashed arrows indicate representative cells on left, similar to \cite{sun_20}.
(\textbf{e}) Evidence cell in Evidence task, similar to \cite{nieh_21}
(\textbf{f}) Cell representation transition during training, illustrating “switch” pattern similar to \cite{zheng_24}.
(\textbf{g}) Classification accuracy of neuron population from dorsal, intermediate and ventral CA1 in CS1234 task. Dashed green line indicates random accuracy. Gray shadow area, cue zone; blue shadow area, action zone. Ventral CA1 cannot distinguish stimulus identity at action zone, which is similar to  \cite{biane_23}.
(\textbf{h}) Dorsal (left), intermediate (middle) and ventral (right) CA1 population representation in CS1234 task. Upper, MDS results shows cues that correspond to the same task outcome (CS1+ and CS2+; CS3- and CS4-) become closer. Lower, representative neuron activity in different cue trial.
(\textbf{i}) Neural manifold approach to task topology during training in Near/far task. 20 trails are grouped as one session. Trials start from the black cross, and go clockwise. Note that the representation gradually “de-correlates” at the action zone (black arrow), resembling a “split-shank wedding ring”, similar to \cite{sun_23}.
} 
\label{fig3}
\end{figure}

To capture the externally and internally driven information, we build a multi-lamellar model (Fig. \ref{fig3}a). The externally driven information (sensory input) acts as the input into the dorsal EC3, and the CA1 readout in ventral lamella would guide the actions. Meanwhile, when $\text{CA1}_i$ in each lamella $i$ reads out the information from $\text{EC3}_i$, and linearly transformed by a matrix $W_i$, $W_i\text{CA1}_i$ would be a part of EC3 input in next lamella $i+1$, jointly influence $\text{EC3}_{i+1}$ with $\text{EC5}_{i+1}$.

The multi-lamellar model performs well in non-trivial tasks (Fig. \ref{fig3}b). Beside for splitter cells and conventional place cells, the proposed model forms lap cell in lap task \cite{sun_20}, evidence cell in evidence task \cite{nieh_21}, delay-active cell (trace cell) in trace task \cite{tanila_97,masuda_20} (Fig. \ref{fig3}c-e).  During training, the cells also undergo representational transition, which is known as tuning changes \cite{sun_23}. Some of the initially silence cells become splitter cells or place cells, and some place cells could also become eventually silence (Fig. \ref{fig3}f). Similar to rodent neurons, the tuning changes could occur rapidly, forming a representation of switching \cite{yang24,zheng_24}. 

Beside for single cell level, our model also exhibits consistency with biological properties in population level, i.e., the population responses are distinct for each trail type. To examine this, we firstly train a linear classifier in order to distinguish the patterns in different trial type. The classifier validation performance, therefore stands for how well the population representation would encode task-relevant information \cite{biane_23}. The model produces similar classification accuracy curve as rodents CA1 cells do (Fig. \ref{fig3}g). 

The multi-lamellar model shows different property along the DV axis, that the representation transits from externally driven in dCA1, to internally driven in vCA1. In task CS1234, the classifier accuracy uncovers this difference (Fig. \ref{fig3}g). In dCA1, the identity of each cue is clearly encoded; while vCA1 distinguishes different cues that can bring different outcome at the action zone, but ignores cue identity, illustrating information abstraction. Additionally, multidimensional scaling (MDS) dimension reduction result on CA1 activity also confirm the difference (Fig. \ref{fig3}h). The above results consist with rodent representation properties \cite{biane_23}. To have an intuitive observation of how the neural representation shapes during learning, we apply a nonlinear dimension reduction technic, uniform manifold approximation and projection (UMAP), on task CS1234 \cite{mcinnes18}. Each population activity is dimensional-reduced to a point in a 3-D space (Fig. \ref{fig3}i). The result shows that, the neural manifold increasingly approaches the task-relevant topology structure during learning, which is similar to physiology results in \cite{sun_23}. 

In short, our multi-lamellar model is capable of conducting detailed observations of the environment, while also abstracting and extracting the underlying task logic based on these observations, which is greatly important in WM and generalization.

\subsection{Learning faster and faster: GATE's capability in generalization and working memory}

\begin{figure}
\centering
\includegraphics[width=1\textwidth]{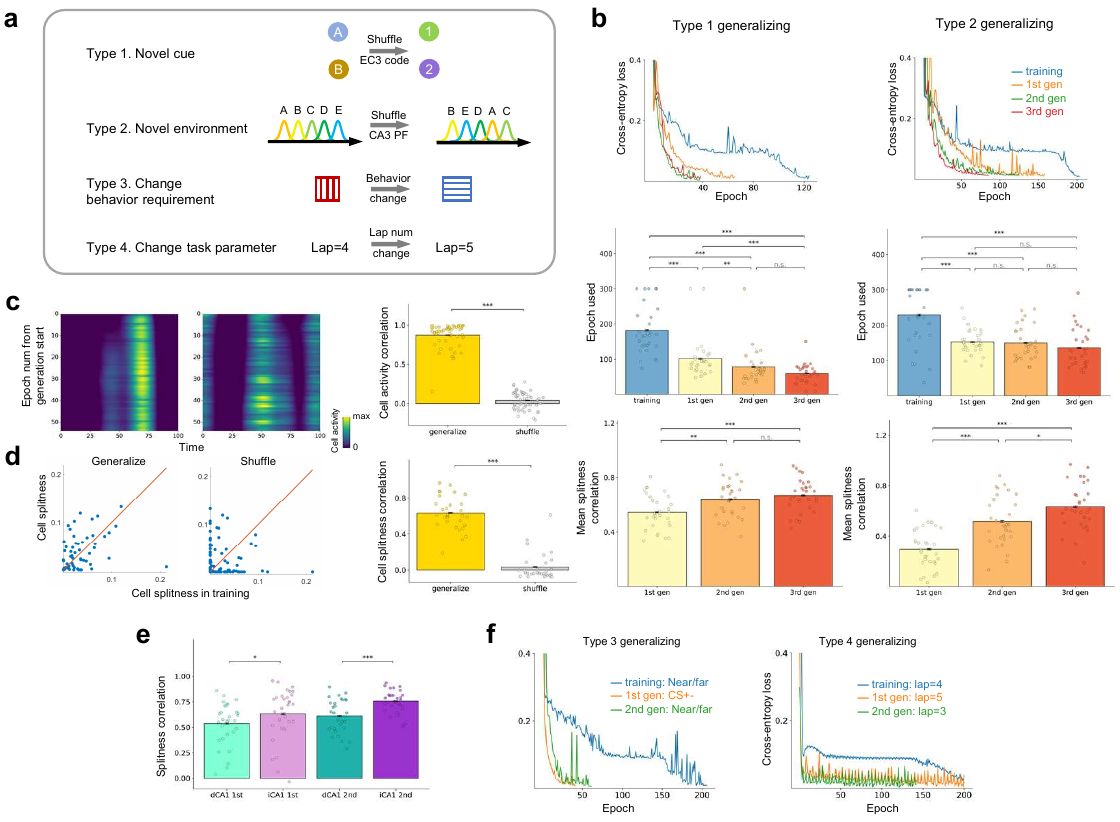}
\caption{
\textbf{Working memory enables generalization.}
(\textbf{a}) Semantic four generalization paradigms. PF, place field.
(\textbf{b}) GATE learns faster and faster during several times of generalization (gen). Left, Type 1 generalization; right, Type 2 generalization. Upper row, representative loss curve (loss over 0.4 is not shown for clarity). Middle, epochs number used when classification loss achieve 0.01, otherwise stops training if epoch exceeds 300. Lower, splitness index correlation between generalization. 
(\textbf{c}) Place field does not significantly change during type 1 generalization. Left, two representative neuron activity. Right, neuron activity correlation during generalization or shuffle. 
(\textbf{d}) Cell representation tend to inherit during generalization. Left, scatter of splitness in training vs. generalization, or training vs. shuffle, red line indicates x=y. Right,  place cell splitness during generalization or shuffle.
(\textbf{e}) splitness correlation between first or second generalization in dCA1 or intermediate CA1 (iCA1), illustrating CA1 splitness differs in distinct lamella. 
(\textbf{f}) Representative loss curve in Type 3 (left) and 4 (right) generalization. In all bar-plots, n=30; two-sided Mann-Whittney U-test; *$P$<0.05, **$P$<0.01, ***$P$<0.001, n.s., not significant.
} 
\label{fig4}
\end{figure}

To assess GATE's adaptability to novel tasks or environments, we modified task settings in four distinct ways (Fig. \ref{fig4}a): (1) replacing the EC3 input with entirely new sensory coding, which corresponds to novel cue types \cite{sun_23}; (2) shuffling all the CA3 place fields (or time fields), simulating a completely new environment; (3) altering the action requirements, such as replacing a CS+- task with Near/far task; and (4) changing task parameters, such as modifying the lap cycle count in the Lap task, while preserving the task's internal logic. Across all conditions, our model adapted to new tasks at an accelerated pace, and exhibited even faster learning in subsequent generalizations (Fig. \ref{fig4}b,f). This mirrors rodent behavior, where learning significantly speeds up after cue replacement with novel sensory inputs \cite{biane_23,sun_23}.

To uncover the basis of our model's adaptability, we analyzed representation changes during generalization. First, we compared place field locations before and after replacing EC3 sensory coding. Most cells with discernible place fields maintained their spatial representations, indicating that generalization involved rate remapping rather than spatial remapping (Fig. \ref{fig4}c). This aligns with findings from \cite{biane_23}, showing that neural representations remain highly similar between old and new tasks after cue replacement, except regions where cues show up.

For environmental generalization, where CA3 fields are manually shuffled, we examine task-relevant representation using a splitness index to quantify each cell's ability to encode distinct task-related information (Fig. \ref{fig4}b,d). Results reveals that splitness remains largely stable, suggesting that task-relevant representations are preserved post-generalization. Notably, intermediate CA1 shows stronger inheritance of these representations than dorsal CA1, underscoring the importance of the DV axis in generalization processes (Fig. \ref{fig4}e).

In conclusion, our findings demonstrate that the model leverages prior experiences by inheriting abstract, task-relevant representations during generalization. When the environment remains unchanged, spatial representations are also retained. These mechanisms enable the model to progressively accelerate its learning, achieving improved performance with generalization.

\section{Discussion}

Working memory and generalization are among the HF's most essential functions and form the foundation of cognitive processes, which stores task-relevant information and rapidly adapts to new environments and tasks. However, the precise mechanisms underlying the integration of working memory and generalization remain uncertain.

We propose GATE framework to address the above issues. In the model, working memory is formalized into two steps: first, information processing -- writing, maintaining, reading, and forgetting; second, information abstracting. Specifically, our model can write, keep and forget information via EC3 persistent activity, selectively readout information by CA1 at appropriate location or time via CA3 gating, and control the working stage of EC3 via EC5 integration. These functions work together to enable flexible selection and integration of new and existing information. Further, our model deploys a multi-lamellar structure, which captures information lamella-wisely.

As a result, the model demonstrated strong performance in more complex tasks. At the level of single-cell representations, our model replicated numerous experimental findings, including the emergence of conventional place cells, splitter cells, lap cells, evidence cells, trace cells, and delay-active cells \cite{masuda_20}. Additionally, at the population level, the model exhibited highly biologically consistent representations. Along the DV axis, dCA1 cells are directly driven by external sensory inputs, whereas vCA1 cells gradually acquire internally driven, composite, and abstract representations through the learning process. Dimensionality reduction methods further reveals how the population encoded task-relevant information, with neural manifolds progressively aligning with task-specific topologies.

Since the working memory in our model can retain both types of disparate information, it exhibits strong adaptability to new environments and tasks while effectively leveraging task-relevant representations acquired from prior learning. Furthermore, generalizing for multiple times across several tasks can enhance the model's working memory capabilities, refining its ability to retain relevant information. This improvement is reflected in progressively faster learning speeds.

Recent studies have revealed a significant correlation between HF activity during information maintenance and performance on human working memory tasks, whereas HF activity during cue presentation shows no such relationship \cite{daume24}. This finding aligns with our model, which posits that the HF sustains activity to retain information about new stimuli for subsequent tasks. Additionally, some studies have demonstrated a positive correlation between persistent activity and working memory load \cite{boran22}. In our model, evidence of information maintenance is reflected in the increased parity consistency and correlation of EC3 activity between cue and reward phases. After training, reward-associated cues also lead to enhanced parity consistency and correlation in CA1, both dorsal and ventral regions. During the maintenance phase, CA1 encodes task-relevant information, with the quantity decreasing over time approximately following an exponential decay pattern \cite{pastalkova_08}. Moreover, EC3 input has been shown to be critical for temporal association memory \cite{suh11}. Persistent firing in lateral entorhinal cortex (LEC) cells correlates with learning ability \cite{lin20}, and lesions in the LEC significantly impair associative memory in rodents \cite{wilson13}. These experimental findings indirectly support our model.

Many existing models of hippocampal WM are built on neural networks with recurrent connections, which employs attractor dynamics or synfire-chains, displaying a capacity to retain information. However, compared to direct persistent activity, feedback network-based models require extensive modifications of feedback weights across different tasks or environments, which limits their generalization ability when facing new tasks or environments. Additionally, rodents with CA3 lesions perform normally in trace tasks, while those with selective CA1 lesions almost completely lose the ability to learn trace associations \cite{manns_05,kesner05}. This phenomenon contradicts the classic hypothesis that WM is maintained through CA3 recurrent networks, but could be explained by the hypothesis that EC3 persistent activity keeps WM. Rodents with CA3 lesions still possess deep CA1 cells, which are primarily driven by EC3 inputs and can read out information stored in EC3 without the need for CA3, leading to more “off-field activity” \cite{sharif21}. In contrast, rodents with CA1 lesions cannot retrieve EC3 information at all.

Many models have attempted to train cognitive map formation within a single task and environment, such as the synfire-chain \cite{wang15}, TEM \cite{whittington_20}, Hebbian-RNN \cite{kappel_14}, CSCG \cite{george_21}, and plasticity model \cite{cone_24}. However, our model goes beyond this by not only explaining how a cognitive map is formed but also demonstrating how it can leverage working memory and past experiences for generalization, allowing rapid adaptation to new environments and tasks. While models like TEM and plasticity model attempt to generalize by replacing part or all of the cues within a task, they fall short when the task's topology changes or when novel situations arise, unlike the flexible cognitive maps our model can produce.

In comparison to gated recurrent neural network (gated RNN) models commonly used in machine learning, such as long-short term memory (LSTM) \cite{hochreiter97} and gated recurrent unit (GRU) \cite{dey17}, our model similarly employs gating mechanisms to control the processes of writing, keeping, reading, and forgetting information. However, gated RNNs struggle to learn tasks where there are long temporal dependencies (e.g., 60–90 steps between sensory input and decision-making) (data not shown). Additionally, at the start of training, these models fail to fully encode external inputs within the network. Despite utilizing latent variables to retain information, they suffer from severe gradient vanishing problems, hindering their performance in such scenarios.

Transformer, recognized as one of the most successful architectures for learning temporal dependencies, offer an alternative approach by encoding states at each time step into vectors and evaluating associations through inner products of linearly transformed vectors. However, for working memory tasks, Transformers require the retention of embeddings at every time step, which is biologically implausible. Moreover, Transformers struggle to form explainable and biologically consistent representations, as highlighted in recent studies \cite{sun_23}.

Our model also offers the following experimentally testable predictions: (1) EC3 possesses information-keeping related neurons, who's on possibility and on time constant correspond to the current task stage; (2) task stages can also be decoded from EC5 activity; (3) learnt behavior co-evolve with vCA1 representation, while lag the dCA1's evolution \cite{cone_24}.

GATE is inspired by numerous biological experimental findings and explains the mechanisms behind certain biological phenomena, and offers some predictions as well. Yet, there are several areas for further study. For instance, some biological mechanisms have not been incorporated into our model. Behavior time-scale synaptic plasticity (BTSP) has been shown to be related to the formation of splitter cells, and due to its strong ability to alter weights, it can rapidly induce changes in CA1 cell representations. Interestingly, though plasticity rule is not deployed in our model, it produces results similar to the switching property of BTSP (Fig. \ref{fig3}g). The hippocampus plays a crucial role in contextual memory, especially in encoding the order of events. Mechanisms like phase precession and SWRs can encode the sequence of events with spike sequences, which might strengthen our model's ability in episodic learning. rodents' HF also shows potential in lifelong learning, which involves learning multiple tasks and rapidly recalling previously learned cognitive map \cite{biane_23}. This process may require involvement from other hippocampal regions, such as the Subiculum \cite{roy17}, Dentate gyrus (DG) \cite{yassa11}, or other areas of the brain.

In summary, GATE effectively captures the HF's working patterns, incorporating both its working memory and generalization abilities. GATE demonstrates that HF, through its three-dimensional multi-lamellar network structure, can form cognitive maps that are transferable across tasks and environments in working memory-related tasks. Furthermore, GATE provides a reliable, explainable and robust brain-inspired framework based on the understanding of HF's mechanism.

\section{Methods}
\label{Methods}
\subsection{External input}

\rowcolors[\hline]{1}{gray!15}{white}  
\begin{table}
\caption{Model parameters}
\centering
\begin{tabular}{p{1.8cm}p{4cm}p{1.8cm}p{7cm}}

Parameter & Value                                                                   & Units     & Description                                                                         \\

$N_{cue}$      & 4 (CS1234 task), 3 (Sequence task), 10 (Trace task), 2 (otherwise) & -         & Number of stimulus cue type in each task                                            \\
$N_{EC3}$      & 100                                                                     & -         & Number of EC3 subgroups                                                   \\
$N_{CA1}$      & 100                                                                     & -         & Number of CA1 neurons                                                               \\
$N_{CA3}$      & 100                                                                     & -         & Number of CA3 neurons                                                               \\
$N_{EC5}$      & 100                                                                     & -         & Number of EC5 neurons                                                               \\
$dt$        & 0.1                                                                     & $dt$ (arb.)    & Time step                                                                   \\
$dx/dt$     & 1                                                                       & $dv$ (arb.) & Velocity of agent in track                                                          \\
$L$         & 100                                                                     & $dx$ (arb.) & Length of the track                                                                 \\
$\Omega_{cue}$  & {[}10, 20]                                                              & $dx$ (arb.) & Cue stimulation zone in track where EC3 subgroups receive cue input           \\
$M_{ij}$       & 0/1 random matrix with $p(1)=0.2$                                         & -         & Cue matrix, which EC3 neurons receive input for each cue type  \\
$c_{01}$, $c_{10}$  & 0.001, 0.02                                                             & -         & Offset of the transit possibility function                                          \\
$h_{01}$, $h_{10}$  & 0.8, 0.6                                                                & -         & Scale coefficient of the transit possibility function                               \\
$m_{01}$, $m_{10}$  & 4, 10                                                                   & -         & Stretch coefficient of the transit possibility function                             \\
$d_{01}$, $d_{10}$  & 1.5, 0.5                                                                & -         & Shift of the transit possibility function                                           \\
$D$         & 5                                                                       & $dx$ (arb.) & Standard deviation of the CA3 place field                                           \\
$C_1$, $C_2$    & 0.2, 1                                                                  & -         & Offset, scale coefficient of the CA1 activation function                            \\
$\epsilon$   & 0.5                                                                     & -         & Small constant to prevent dividing zero and filter out inferior neurons             \\
$bs$        & 32, 128 (Sequence task and Lap task only)                               & -         & Batch size of the training samples                                                  \\
$lr$        & 0.01                                                                    & -         & Learning rate of the Adam optimizer   

\end{tabular}
\label{tab:param_table}
\end{table}

Parameters of all methods are listed in Tab. ~\ref{tab:param_table}. For simplicity, the agent runs with a constant unit velocity through the whole track, such that $x=t$. 

External sensory input drives the EC3 neurons in first lamella via a short pulse. When a task has $N_{cue}$ types of cue, and the $j$-th cue type is deployed in a training trial, the cue stimulates several EC3 neurons as follow: 
\begin{equation}
    cue_i (t) =\chi (t)  M_{ij}
\end{equation}
where M is a 0-1 matrix, randomly defined before training so that each cue stimulates specific EC3 neurons; $\chi$ is an indicator function that limit the cue to show up in cue zone $\Omega_{cue}$:
\rowcolors{4}{white}{white}
\begin{equation}
    \chi(t) = \begin{cases} 
      1 & \text{if } t\in\Omega_{cue} \\
      0 & \text{otherwise}
   \end{cases}
\end{equation}

\subsection{EC3 population model}

A single EC3 Markov neuron transits its state base on its current state and its input. Given $\text{EC3}_{input}$ (see below), the possibility of transition from on to off is $p_{10}=\gamma_{10} (\text{EC3}_{input})$, transit from off to on with possibility $p_{01}=\gamma_{01}(\text{EC3}_{input})$, where $\gamma_{10}$, $\gamma_{01}$ are non-linear functions: 
\begin{equation} 
    \gamma_f(x) = c_f + h_f \sigma\left[m_f(x-d_f)\right]
\end{equation} 
in which $\sigma(x)$ is the sigmoid function: $\sigma(x)=1/(1+e^{-x})$, $f=(01)$ or $f=(10)$ . Therefore, on a population level, the proportion of "on" cell in $i$-th EC3 subgroup $r_i$, follows an ODE: 
\begin{equation}
    \frac{dr_i}{dt} = (1-r_i)  p_{01_i} + r_i  (1-p_{10_i})
\end{equation}

When given a static $\text{EC3}_{input}$, this ODE has a global asymptotic stable point, $r_\infty=p_{01}/(p_{01}+ p_{10})$, and approach to $h_\infty$ with a time constant, $\tau=1/(p_{01}+p_{10})$. The stage of the EC3 subgroup can be identified by these two parameters, including writing, keeping and forgetting. In doing so, EC3 subgroups in our model can be discretized into single EC3 Markov chain after training (each subgroup is replaced by ten distinct Markov chains, but with the same EC3 input).

\subsection{Hippocampus formation network}

The output of the $m$-th neuron in CA3 is modeled as a Gaussian-like function on time $t$: 
\begin{equation}
    g_m(t)=exp(- (t-t_m)^2/D^2)
\end{equation}
where $D$ stands for the standard deviation, $t_m$ stands for the center of the place field that covers the whole track. For simplicity, the agent moves on a conveyor belt with a constant speed, i.e., the track is connected at the beginning and end, forcing the agent to forget the information in the last trial.

The $j$-th CA1 neuron can be described by the following multi-compartment model: 

\begin{equation}
    \begin{aligned}
    b_j(t) &= \text{relu}\left(W_{jm}^{basal} g_m(t)\right) \\
    a_j(t) &= \sigma\left(W_{ji}^{apical} r_i(t) - \alpha_j \right) \\
    s_j(t) &= \text{relu}\left(b_j(t) * (C_1+C_2a_j(t)) - \beta_j \right)
    \end{aligned}
\end{equation}

where $b_j$ is the basal potential, $a_j$ is the apical potential, $W^{basal}$ is the basal weight, $W^{apical}$ is the apical weight, and $\alpha_j$, $\beta_j$ are learnable inhibitory biases, $C_1$, $C_2$ are constants. When basal potential is weak, the CA1 neuron output is close to zeros, i.e., CA3 input ‘gates' the CA1 readout. When $b_j$ is strong enough, $a_j$ would act like a ‘gain factor' \cite{larkum04}: potentiate the CA1 output when $a_j$ is large, while depress the CA1 output when $a_j$ is small. These cellular mechanisms guarantee that the CA1 neurons can learn a complicated cognitive map, including both conventional place field and task-relevant cell. 

The $k$-th EC5 neuron integrates its CA1 input: 
\begin{equation}
    v_k (t)=\text{clip}\left(\int_0^t\phi(W_{kj}^{EC5} s_j (x))dx\right)
\end{equation}
where $W^{EC5}$ is the weight initiated as an identical matrix, $\phi$ is a threshold function that ignores small input, and $\text{clip}$ function is applied to limit output range of EC5 to $[-1, 1]$. EC5 output, would be subsequently transmitted into EC3 input of the same lamella:
\begin{equation}
    EC3_{input_i}(t)=W_{ik}^{EC5} v_k (t)+cue_i (t)+W_{ij}^{DV} \tilde{s_j} (t)
\end{equation}
where $W^{EC3}$ is the weight, $cue_i$ is the sensory input, $W^{DV}$ is the dorsoventral transition weight, and $\tilde{s_j}$ is the CA1 output in the previous lamella. The second term is only included when the EC3 neuron belongs to the dorsal lamella, while the third term is only included in other lamellae.

\subsection{Agent behavior and training}

Agent behavior is derived from the CA1 output $\hat{s_j}$ in ventral lamella through a policy matrix $W^{action}$:
\begin{equation}
    q_n(t)=W_{nj}^{action} \hat{s_j} (t)
\end{equation}
where $n$ indexes over possible actions, and action with higher $q$ will be chosen. In our tasks, the agent needs to determine whether to ‘lick the feeding tube' at each time steps \cite{biane_23}, which forms a binary classification paradigm. When ‘reward is provided via feeding tube', the agent should lick the tube, and vice versa. With the label given, our model can be trained with back propagation, with a cross-entropy loss with classification weight $W_{lick}$, and Adam optimizer with learning rate $lr$. Further, the data is batch-normalized to accelerate training with batch size $bs$. When generalizing, $W^{action}$ is reset while other weights are retained.

\subsection{Splitness index}

The splitness index, $S_j$of the $j$-th cell in CA1, is defined as follow: 
\begin{equation}
    S_j = \frac{\text{std}_l \left( max_t\left[\bar{s}_j^l(t) \right]\right) }  {\text{mean}_l \left( max_t \left[ \bar{s}_j^l(t) \right]\right) + \epsilon}
\end{equation}
where $\bar{s}^l_j$ is the mean neuron output in trials with cue type $l$, and $\epsilon$ is a small constant to prevent dividing zero, and filter out inferior neurons with low activity.


\section{Code Availability}
All simulations and training are run via custom code in Python 3.10.11 / Pytorch 2.0.1. The code will be available when the paper is accepted. 

\section{Acknowledgments}
This research was supported by the National Natural Science Foundation of China (Nos. 12271429, 12090021, and 12226007).

\section{Author Contributions}
Yuechen Liu: Conceptualization, methodology, investigation, formal analysis, data curation, writing -- original draft, writing -- review and editing; Zishun Wang: Formal analysis, writing -- review and editing; Zongben Xu: Methodology, supervision, funding acquisition, writing -- review and editing; Chen Qiao: Methodology, supervision, funding acquisition,  project administration, writing -- original draft, writing -- review and editing.

\section{Competing interests}
The authors declare no competing interests.

\bibliographystyle{unsrt}  

\begin{thebibliography}{1}  


	\bibitem{baddeley12} Baddeley, A. Working memory: Theories, models, and controversies. {\em Annual Review Of Psychology}. \textbf{63}, 1-29 (2012)

	\bibitem{banich11} Banich, M. \& Caccamise, D. Generalization of knowledge: Multidisciplinary perspectives. (Psychology Press, Hove,2011)

	\bibitem{daume24} Daume, J., Kamiński, J., Salimpour, Y., Schjetnan, A., Anderson, W., Valiante, T., Mamelak, A. \& Rutishauser, U. Persistent activity during working memory maintenance predicts long-term memory formation in the human hippocampus. {\em Neuron}. \textbf{112}, 3957-3968. e3 (2024)

	\bibitem{biane_23} Biane, J., Ladow, M., Stefanini, F., Boddu, S., Fan, A., Hassan, S., Dundar, N., Apodaca-Montano, D., Zhou, L. \& Fayner, V. Neural dynamics underlying associative learning in the dorsal and ventral hippocampus. {\em Nature Neuroscience}. \textbf{26}, 798-809 (2023)

	\bibitem{pastalkova_08} Pastalkova, E., Itskov, V., Amarasingham, A. \& Buzsaki, G. Internally generated cell assembly sequences in the rat hippocampus. {\em Science}. \textbf{321}, 1322-1327 (2008)

	\bibitem{sun_23} Sun, W., Winnubst, J., Natrajan, M., Lai, C., Kajikawa, K., Michaelos, M., Gattoni, R., Stringer, C., Flickinger, D. \& Fitzgerald, J. Learning produces a hippocampal cognitive map in the form of an orthogonalized state machine. {\em BioRxiv}. pp. 2023.08. 03.551900 (2023)

	\bibitem{wang15} Wang, Y., Romani, S., Lustig, B., Leonardo, A. \& Pastalkova, E. Theta sequences are essential for internally generated hippocampal firing fields. {\em Nat Neurosci}. \textbf{18}, 282-8 (2015)

	\bibitem{fortin_02} Fortin, N., Agster, K. \& Eichenbaum, H. Critical role of the hippocampus in memory for sequences of events. {\em Nature Neuroscience}. \textbf{5}, 458-462 (2002)

	\bibitem{nieh_21} Nieh, E., Schottdorf, M., Freeman, N., Low, R., Lewallen, S., Koay, S., Pinto, L., Gauthier, J., Brody, C. \& Tank, D. Geometry of abstract learned knowledge in the hippocampus. {\em Nature}. \textbf{595}, 80-84 (2021)

	\bibitem{sun_20} Sun, C., Yang, W., Martin, J. \& Tonegawa, S. Hippocampal neurons represent events as transferable units of experience. {\em Nature Neuroscience}. \textbf{23}, 651-663 (2020)

	\bibitem{macdonald_11} MacDonald, C., Lepage, K., Eden, U. \& Eichenbaum, H. Hippocampal "time cells" bridge the gap in memory for discontiguous events. {\em Neuron}. \textbf{71}, 737-49 (2011)

	\bibitem{mcechron_97} McEchron, M. \& Disterhoft, J. Sequence of single neuron changes in CA1 hippocampus of rabbits during acquisition of trace eyeblink conditioned responses. {\em Journal Of Neurophysiology}. \textbf{78}, 1030-1044 (1997)

	\bibitem{tahvildari_07} Tahvildari, B., Fransén, E., Alonso, A. \& Hasselmo, M. Switching between “On” and “Off” states of persistent activity in lateral entorhinal layer III neurons. {\em Hippocampus}. \textbf{17}, 257-263 (2007)

	\bibitem{jochems13} Jochems, A., Reboreda, A., Hasselmo, M. \& Yoshida, M. Cholinergic receptor activation supports persistent firing in layer III neurons in the medial entorhinal cortex. {\em Behav Brain Res}. \textbf{254} pp. 108-15 (2013)

	\bibitem{grienberger_22} Grienberger, C. \& Magee, J. Entorhinal cortex directs learning-related changes in CA1 representations. {\em Nature}. \textbf{611}, 554-562 (2022)

	\bibitem{koster_18} Koster, R., Chadwick, M., Chen, Y., Berron, D., Banino, A., Düzel, E., Hassabis, D. \& Kumaran, D. Big-loop recurrence within the hippocampal system supports integration of information across episodes. {\em Neuron}. \textbf{99}, 1342-1354. e6 (2018)

	\bibitem{rubin14} Rubin, R., Watson, P., Duff, M. \& Cohen, N. The role of the hippocampus in flexible cognition and social behavior. {\em Front Hum Neurosci}. \textbf{8} pp. 742 (2014)

	\bibitem{tanila_97} Tanila, H., Shapiro, M. \& Eichenbaum, H. Discordance of spatial representation in ensembles of hippocampal place cells. {\em Hippocampus}. \textbf{7}, 613-623 (1997)

	\bibitem{colgin08} Colgin, L., Moser, E. \& Moser, M. Understanding memory through hippocampal remapping. {\em Trends Neurosci}. \textbf{31}, 469-77 (2008)

	\bibitem{fanselow10} Fanselow, M. \& Dong, H. Are the dorsal and ventral hippocampus functionally distinct structures?. {\em Neuron}. \textbf{65}, 7-19 (2010)

	\bibitem{whittington_20} Whittington, J., Muller, T., Mark, S., Chen, G., Barry, C., Burgess, N. \& Behrens, T. The Tolman-Eichenbaum machine: unifying space and relational memory through generalization in the hippocampal formation. {\em Cell}. \textbf{183}, 1249-1263. e23 (2020)

	\bibitem{kappel_14} Kappel, D., Nessler, B. \& Maass, W. STDP installs in Winner-Take-All circuits an online approximation to hidden Markov model learning. {\em PLoS Comput Biol}. \textbf{10}, e1003511 (2014)

	\bibitem{george_21} George, D., Rikhye, R., Gothoskar, N., Guntupalli, J., Dedieu, A. \& Lázaro-Gredilla, M. Clone-structured graph representations enable flexible learning and vicarious evaluation of cognitive maps. {\em Nature Communications}. \textbf{12}, 2392 (2021)

	\bibitem{cone_24} Cone, I. \& Clopath, C. Latent representations in hippocampal network model co-evolve with behavioral exploration of task structure. {\em Nature Communications}. \textbf{15}, 687 (2024)

	\bibitem{ainge_07} Ainge, J., Van Der Meer, M., Langston, R. \& Wood, E. Exploring the role of context‐dependent hippocampal activity in spatial alternation behavior. {\em Hippocampus}. \textbf{17}, 988-1002 (2007)

	\bibitem{zhao_22} Zhao, X., Hsu, C. \& Spruston, N. Rapid synaptic plasticity contributes to a learned conjunctive code of position and choice-related information in the hippocampus. {\em Neuron}. \textbf{110}, 96-108 e4 (2022)

	\bibitem{kinkhabwala_20} Kinkhabwala, A., Gu, Y., Aronov, D. \& Tank, D. Visual cue-related activity of cells in the medial entorhinal cortex during navigation in virtual reality. {\em Elife}. \textbf{9} pp. e43140 (2020)

	\bibitem{jarsky_05} Jarsky, T., Roxin, A., Kath, W. \& Spruston, N. Conditional dendritic spike propagation following distal synaptic activation of hippocampal CA1 pyramidal neurons. {\em Nat Neurosci}. \textbf{8}, 1667-76 (2005)

	\bibitem{salz_16} Salz, D., Tiganj, Z., Khasnabish, S., Kohley, A., Sheehan, D., Howard, M. \& Eichenbaum, H. Time cells in hippocampal area CA3. {\em Journal Of Neuroscience}. \textbf{36}, 7476-7484 (2016)

	\bibitem{manns_05} Manns, J. \& Eichenbaum, H. Time and treason to the trisynaptic teachings: theoretical comment on Kesner et Al. (2005). {\em Behav Neurosci}. \textbf{119}, 1140-3 (2005)

	\bibitem{egorov_02} Egorov, A., Hamam, B., Fransén, E., Hasselmo, M. \& Alonso, A. Graded persistent activity in entorhinal cortex neurons. {\em Nature}. \textbf{420}, 173-178 (2002)

	\bibitem{masuda_20} Masuda, A., Sano, C., Zhang, Q., Goto, H., McHugh, T., Fujisawa, S. \& Itohara, S. The hippocampus encodes delay and value information during delay-discounting decision making. {\em Elife}. \textbf{9} (2020)

	\bibitem{yang24} Yang, W., Sun, C., Huszár, R., Hainmueller, T., Kiselev, K. \& Buzsáki, G. Selection of experience for memory by hippocampal sharp wave ripples. {\em Science}. \textbf{383}, 1478-1483 (2024)

	\bibitem{zheng_24} Zheng, Z., Huszar, R., Hainmueller, T., Bartos, M., Williams, A. \& Buzsaki, G. Perpetual step-like restructuring of hippocampal circuit dynamics. {\em Cell Rep}. \textbf{43}, 114702 (2024)

	\bibitem{mcinnes18} McInnes, L., Healy, J. \& Melville, J. Umap: Uniform manifold approximation and projection for dimension reduction. {\em ArXiv Preprint}. (2018), https://arxiv.org/abs/1802.03426v2

	\bibitem{boran22} Boran, E., Hilfiker, P., Stieglitz, L., Sarnthein, J. \& Klaver, P. Persistent neuronal firing in the medial temporal lobe supports performance and workload of visual working memory in humans. {\em Neuroimage}. \textbf{254} pp. 119123 (2022)

	\bibitem{suh11} Suh, J., Rivest, A., Nakashiba, T., Tominaga, T. \& Tonegawa, S. Entorhinal cortex layer III input to the hippocampus is crucial for temporal association memory. {\em Science}. \textbf{334}, 1415-20 (2011)

	\bibitem{lin20} Lin, C., Sherathiya, V., Oh, M. \& Disterhoft, J. Persistent firing in LEC III neurons is differentially modulated by learning and aging. {\em Elife}. \textbf{9} pp. e56816 (2020)

	\bibitem{wilson13} Wilson, D., Watanabe, S., Milner, H. \& Ainge, J. Lateral entorhinal cortex is necessary for associative but not nonassociative recognition memory. {\em Hippocampus}. \textbf{23}, 1280-1290 (2013)

	\bibitem{kesner05} Kesner, R., Hunsaker, M. \& Gilbert, P. The role of CA1 in the acquisition of an object-trace-odor paired associate task. {\em Behav Neurosci}. \textbf{119}, 781-6 (2005)

	\bibitem{sharif21} Sharif, F., Tayebi, B., Buzsaki, G., Royer, S. \& Fernandez-Ruiz, A. Subcircuits of Deep and Superficial CA1 Place Cells Support Efficient Spatial Coding across Heterogeneous Environments. {\em Neuron}. \textbf{109}, 363-376 e6 (2021)

	\bibitem{hochreiter97} Hochreiter, S. Long Short-term Memory. {\em Neural Computation MIT-Press}. (1997)

	\bibitem{dey17} Dey, R. \& Salem, F. Gate-variants of gated recurrent unit (GRU) neural networks. {\em 2017 IEEE 60th International Midwest Symposium On Circuits And Systems (MWSCAS)}. pp. 1597-1600

	\bibitem{roy17} Roy, D., Kitamura, T., Okuyama, T., Ogawa, S., Sun, C., Obata, Y., Yoshiki, A. \& Tonegawa, S. Distinct Neural Circuits for the Formation and Retrieval of Episodic Memories. {\em Cell}. \textbf{170}, 1000-1012 e19 (2017)

	\bibitem{yassa11} Yassa, M. \& Stark, C. Pattern separation in the hippocampus. {\em Trends Neurosci}. \textbf{34}, 515-25 (2011)

	\bibitem{larkum04} Larkum, M., Senn, W. \& Luscher, H. Top-down dendritic input increases the gain of layer 5 pyramidal neurons. {\em Cereb Cortex}. \textbf{14}, 1059-70 (2004)



\end{thebibliography}


\end{document}